\documentclass{jfm}
\usepackage{graphicx}
\usepackage{subfigure,rotating,amsmath,float,upmath,caption}
\usepackage{IEEEtrantools}
\usepackage{natbib}
\usepackage{color}
\ifCUPmtlplainloaded \else
  \checkfont{eurm10}
  \iffontfound
    \IfFileExists{upmath.sty}
      {\typeout{^^JFound AMS Euler Roman fonts on the system,
                   using the 'upmath' package.^^J}%
       \usepackage{upmath}}
      {\typeout{^^JFound AMS Euler Roman fonts on the system, but you
                   dont seem to have the}%
       \typeout{'upmath' package installed. JFM.cls can take advantage
                 of these fonts,^^Jif you use 'upmath' package.^^J}%
      }
  \else
  \fi
\fi

\ifCUPmtlplainloaded \else
  \checkfont{msam10}
  \iffontfound
    \IfFileExists{amssymb.sty}
      {\typeout{^^JFound AMS Symbol fonts on the system, using the
                'amssymb' package.^^J}%
       \usepackage{amssymb}%

      }{}
  \fi
\fi

\ifCUPmtlplainloaded \else
  \IfFileExists{amsbsy.sty}
    {\typeout{^^JFound the 'amsbsy' package on the system, using it.^^J}%
     \usepackage{amsbsy}}
    {\providecommand\boldsymbol[1]{\mbox{\boldmath $##1$}}}
\fi

\def \t  {\theta}

{}

\def \Ca  {\mbox{Ca}}

\def \rhod {\rho_{2}}
\def \rhoc {\rho_{1}}
\def \mud {\mu_{2}}
\def \muc {\mu_{1}}
\def \fn  {f_{n}}

\def \Pn  {P_{n}}
\def \bft{{\mathbf{f} \lp T \rp}}

\def \bp {\mathbf{p}}

\def \F {\mathbf{F}}
\def \Rthe {R \lp  \t \rp}

\def \gt  {G^{\mathrm{max}}\lp T \rp}

\def \gdelsmax {G_{\delS}^{\mathrm{max}}}

\def \bn {\mathbf{n}}
\def \bu {\mathbf{u}}
\def \bue {\bu_{1}}
\def \bui {\bu_{2}}
\def \bsig {\boldsymbol{\sigma}}
\def \uter {U_{\mathrm{ter}}}
\def \d    {\mathrm{d}}

\def \Cacri {\Ca_{\mathrm{c}}}

\def \delS {\Delta S}
\def \Marea {\mathbf{M}_{\delS}}
\def \gtareamax {G^{\mathrm{max}}_{\delS}\lp T \rp}
\def \gtarea {G_{\delS}\lp t \rp}
\def \gareamax {G^{\mathrm{max}}_{\delS}}
\def \garea {G_{\delS}}

\def \delcri {\delta_{\mathrm{c}}}

\def \tmax {T_{\mathrm{max}}}
\def \f {\mathbf{f}}

\def \fo {{\f}^{\mathrm{opt}}_{\left[T\right]}}
\def \fomax {{\f}^{\mathrm{opt}}_{\left[\tmax\right]}}

\def \foi {\fo \lp 0 \rp}

\def\F{{\mathbf{F}}}

\def\acca{{\mathbf{h}}}
\def\T{{\mathbf{T}}}

\def\varepsilon{\epsilon}

\def\A{\stackrel{\triangledown}{\mathbf{A}}}

\def \lp  {\left(}
\def \rp  {\right)}

\def \A   {\mathbf{A}}

\title[Inertialess non-modal energy growth of a rising droplet]{The stability of 
a rising 
droplet: an inertialess non-modal growth mechanism}

\author[G. Gallino and L. Zhu and F. Gallaire]%
{Giacomo Gallino, Lailai Zhu and Fran\c{c}ois Gallaire$^{1}$%
  \thanks{Email address for correspondence: francois.gallaire@epfl.ch}}

\affiliation{$^1$Laboratory of Fluid Mechanics and Instabilities, \'{E}cole Polytechnique F\'{e}d\'{e}rale de Lausanne, Lausanne, CH-1015, Switzerland}

\begin{document}
\maketitle 

\begin{abstract}
Prior modal stability analysis~\citep{kojima1984formation} predicted that a 
rising or sedimenting droplet in a viscous fluid is stable in the presence of surface tension no matter how small, in 
contrast to 
experimental and numerical results. By performing a non-modal stability 
analysis, 
we demonstrate the potential for 
transient growth of the interfacial energy of a rising droplet in the limit of 
inertialess Stokes equations.
The predicted critical capillary numbers for transient growth agree well with those
for unstable shape evolution of droplets found in the 
direct numerical simulations of ~\citet{koh1989stability}.
Boundary integral simulations are used to 
delineate the critical amplitude of the most destabilizing perturbations. 
The critical amplitude is negatively correlated with the linear optimal energy 
growth, implying that the transient growth is 
responsible for reducing the necessary perturbation amplitude required to 
escape the basin of attraction of the spherical solution.

\end{abstract}

\section{Introduction}\label{sec:introduction}
The instability of capillary interfaces has long been an intriguing topic in fluid mechanics. Perhaps one of the earliest 
investigated interfacial instability phenomena is the Rayleigh-Taylor 
instability, where a denser fluid  located above  a
lighter one protrudes into the latter due to any arbitrary small 
perturbation of the initially flat interface.
However, this protrusion is not always observed when a droplet rises or 
sediments into another density-contrasted fluid. According to 
~\citet{hadamard} and ~\citet{rybzynski}, a spherical translating droplet is a solution of 
this problem in the Stokes regime, regardless of the presence or magnitude of 
the surface tension. What remains unknown, however, is the existence of other 
equilibrium shapes of the droplet and the influence of surface tension on 
the stability of the spherical solution.

Experiments were conducted by ~\citet{kojima1984formation} to examine this issue.
Two patterns of shape instability were observed: depending on the viscosity 
ratio $\lambda$, a protrusion or an indentation at the rear of droplet was seen 
to grow with time. 
~\citet{kojima1984formation} also performed a linear stability analysis assuming that the droplet underwent small
deformations. A linear operator depending on the viscosity ratio $\lambda$ and capillary number $\Ca$ (inversely 
scaling with the surface tension) was derived which governs
the linearised droplet shape evolution. 
It was found that, irrespective of the value of $\Ca$, i.e. even for arbitrary 
small surface tension, the eigenvalues of the operator had negative real part, 
pointing to a linearly stable shape. 
The authors recognized that this linear stability study contradicted their experiments showing instabilities with finite surface tension; 
Direct numerical simulations (DNS)~\citep{koh1989stability} also reported the 
unstable shape evolution of slightly disturbed droplets in the presence of sufficient surface tension ($\Ca < 
10$). Recent numerical work has examined the effect of surfactants~\citep{johnson2000stability} and 
viscoelasticity~\citep{wu2012stability} on this scenario.

The contradiction between the theory and experiments/DNS is somewhat reminiscent of the case of the fingering instability of a film flowing
down an inclined plane: the 
experimentally-measured~\citep{huppert1982flow,de1992growth} critical 
inclination angle 
triggering instability was found to be well below that obtained from the linear theory. ~\citet{bertozzi1997linear} discovered that 
the traditional spectrum analysis failed to capture the short-time but significant energy amplification of the 
perturbations near the contact line. They pinpointed the missing mechanism by performing a so-called 
non-modal analysis, borrowed from the transient growth theory founded and developed in the 
early $1990$s for hydrodynamic stability analysis~\citep{trefethen1993hydrodynamic,reddy1993energy, baggett1995mostly}, to identify  and interpret the short-time energy amplification.

The non-modal tools of stability theory have been used to explain the discrepancies 
between the theoretically computed critical Reynolds number and the experimentally-observed  counterpart in a variety 
of wall-bounded shear flows~\citep{schmid2007nonmodal}. The traditional eigenvalue analysis as also used in 
~\citet{kojima1984formation}, i.e. the so-called modal approach, can sometimes 
fail 
to interpret real flow dynamics as the spectrum of the linear operator only dictates the 
asymptotic fate of the perturbations \textit{without} considering their \textit{short-term} 
dynamics~\citep{schmid2001stability}. The non-modal analysis, in contrast, is able to capture the short-time perturbation 
characteristics and determine the most dangerous initial conditions leading to the optimal energy growth. In 
addition to its great success in the traditional hydrodynamic stability analysis, it has been also used 
to elucidate complex flow instability problems including capillary interfaces~\citep{davis2003generalized}, 
thermal-acoustic interactions~\citep{balasubramanian2008thermoacoustic, juniper2011triggering} and 
viscoelasticity~\citep{jovanovic2010transient}.

In this paper, we perform a non-modal analysis to investigate the shape 
instability of a rising droplet in an ambient fluid, neglecting 
inertial effects. After introducing the linearised equations and operator in 
Sec.~\ref{sec:problem} and
the non-modal approaches in Sec.~\ref{sec:methods}, we demonstrate the existence 
of transient growth and predict the
critical capillary numbers required for instability to become possible in 
Sec.~\ref{sec:linear_nonmodal}. In Sec.~\ref{sec:lin_non_evol}, we conduct 
in-house DNS to compute the nonlinear shape evolutions of the droplets 
initiated 
with the linear optimal perturbations and identify the minimal
amplitudes leading eventually to instability. We further analyse the 
relationship 
between the 
optimal growth and the critical amplitude of perturbation. We finally examine
how the instability pattern is related to the viscosity ratio and propose a phenomenological explanation in Sec.~\ref{sec:conclusions}.

\section{Governing equations and linearisation}\label{sec:problem}
We study the dynamics of a buoyant droplet rising in an ambient fluid in the Stokes regime.
The droplet is assumed to be axisymmetric and the axis is along the $z$ direction with gravity ${\bf g}=-g {\bf e}_z$.
The two Newtonian immiscible fluids, one carrying the droplet (fluid $2$), and the other constituting the droplet (fluid $1$) are characterized by different densities $\rho_{2}>\rho_{1}$, inducing (without loss of generality) an upward migration of the droplet. Likewise, their viscosities are 
$\mu_2$ and $\mu_1$ respectively, with a ratio $\lambda=\mu_1/\mu_2$. The interface between the two fluids 
has a uniform and constant surface tension coefficient $\gamma$. The undeformed state of the droplet is a sphere of 
radius $a$ and terminal velocity
$\uter = \frac{a^{2}g (\rho_{2} - \rho_{1}) }{\mu_2} \frac{1+\lambda}{3\lp 1+
3\lambda/2 \rp}$ ~\citep{leal2007advanced}. 
We use $a$ and $\uter$ as the reference length and velocity scales, and
$\mu_2 \uter/a$ as the reference scale
for $p$ and $\bsig$, the modified pressure (removing the hydrostatic part) and the corresponding 
stress tensor respectively~\citep{batchelor2000introduction}. Hence, the governing equations for the non-dimensional velocity and pressure field inside the droplet $\left( \mathbf{u}_1, p_1 \right)$ and that outside the droplet $\left( \mathbf{u}_2, p_2 \right)$ are written as
\begin{align}\label{equ:stokes}
& \nabla \cdot \bu_1 = 0, -\nabla p_{1} +\lambda \nabla^{2} \bu_1 = 0, \nonumber \\
& \nabla \cdot \bu_2 = 0, -\nabla p_{2} + \nabla^{2} \bu_2 = 0,
\end{align}
where the velocity is zero at infinity and
the boundary conditions on the
interface are
\begin{align}\label{equ:bcinter}
 \bue &= \bui, \nonumber \\
 \bsig_{2} \cdot \bn - \bsig_{1} \cdot \bn &= \left[\nabla \cdot \bn /\Ca + 3 z \lp 1 + 3\lambda/2 
\rp/\lp 1+\lambda \rp \right] \bn.
\end{align}
Here, $\bn$ is the unit normal vector pointing from the interface towards the 
carrier fluid and $\Ca=\mu_2\uter/\gamma$ is the capillary number indicating the ratio of the viscous effect
with respect to the surface tension effect.

\begin{figure}
\centering
	\includegraphics[scale=0.5]{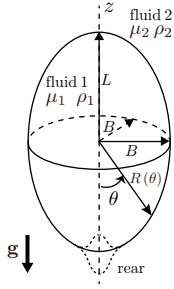}
		\caption{An axisymmetric droplet rising in the quiescent 
fluid, along the axial ($z$) 
direction. The fluid inside and outside is labelled as fluid 1 and fluid 2 
respectively, so as their dynamic
viscosities ($\muc$, $\mud$) and densities ($\rhoc$, $\rhod$). The polar 
coordinates $R\lp \theta \rp$ are used to 
represent its shape, where $\theta$ is measured from its rear stagnation point. $L$ and $B$ is the axis 
length along the 
revolution axis and orthogonal directions.}  
\label{fig:sketch}
\end{figure}
Following~\citet{kojima1984formation}, the interface of an axisymmetric droplet undergoing small deformation
can be expressed in polar coordinates as
 \begin{equation}\label{equ:R_drop}
  R \lp \t \rp = 1 + \delta \sum\nolimits_{n=2}^{\infty}\lp 2n+1 \rp \fn \Pn \lp \cos{\t}\rp, 
 \end{equation}
 where $\t$ is the polar angle measured from the rear of droplet, $\Rthe$ is the polar distance (see 
figure~\ref{fig:sketch}), $\delta$ indicates the amplitude of the deformation, the 
$\Pn$ are the $n$th-order Legendre 
polynomials and the $\fn$ are the corresponding coefficients. The first two
terms $P_{0}$ and $P_{1}$  are removed such that the volume of the droplet is conserved and its centroid stays 
at the origin~\citep{kojima1984formation}.
To advance the interface, the kinematic condition $\partial R(\theta,t)/\partial t = \bu (\theta,t) \cdot \bn$ is applied.

Following ~\citet{kojima1984formation} and linearising the governing equations and truncating the series expansion, the 
evolution of the droplet can be obtained by solving a system of ordinary differential equations,
\begin{equation}\label{equ:linear_evol}
\d \f/\d t = \A \f,
\end{equation} 
where the shape coefficient $\f=\lp f_{2},f_{3},...,f_{m+1} \rp^{\mathrm{T}}$ is a truncated vector and 
$\A$ is an $m \times m$ matrix depending on $\lambda$ and $\Ca$.
It should be noted that the shape of the droplet can be expressed by a unique series of 
coefficients $\f \delta$ and vice 
versa; for a certain $\f$, the effective shape varies significantly with the 
amplitude and sign of $\delta$. For the truncation of $\f$ we use $m=1000$ 
throughout our study: extensive tests using larger values of $m$
confirm that our results are independent of this truncation level.

\section{Non-modal analysis: Theory}\label{sec:methods}
As shown by the modal analysis of ~\citet{kojima1984formation}, the operator 
$\A$ has a stable spectrum with all of its 
eigenvalues having negative real parts, irrespective of the magnitude of the 
surface tension, as long as the capillary number is finite. This model analysis
predicts the long-term behaviour of the disturbance but in the short-term limit it is only valid
if the linear operator $\A$ is normal, i.e.
its eigenvectors are orthogonal. In the case of a non-normal operator, even
though the amplitudes of all eigenmodes 
decay exponentially, their nonorthogonality can lead to a transient energy growth over a 
short time. We indeed found that $\A$ was non-normal. 
The  optimal growth 
$G^{\mathrm{max}}$ of the initial energy  ($L_2$ norm) over a chosen time interval $ [0,T]$~\citep{schmid2007nonmodal} is
\begin{align}\label{equ:gtniv}
\gt &= \max_{\f \lp 0 \rp} \left[ G\lp T \rp = \frac{||\bft||_{2}}{||\f \lp 0 \rp||_{2}} \right]  = ||\exp{\lp T\A 
\rp}||_{2},
\end{align}
where $\f \lp 0 \rp$ denotes the initial perturbation. $\gt$ represents the 
maximum amplification of the 
initial energy at a target time (the so-called horizon) $T$ where the 
optimization has been performed over all possible perturbations $\f \lp 0 \rp$. 
The  optimal 
initial perturbation for horizon $T$ will be denoted $\foi$.
The quantity $G^{\mathrm{max}}$ is the envelope of all individual gain profiles, 
indicating the presence of transient growth when $G^{\mathrm{max}}\lp 
T \rp>1$ for some $T$.

Compared with the $L_2$ norm in equ.~\ref{equ:gtniv}, it is natural to introduce a physically-driven form of energy, 
designed for the physical problem at hand. In the present study,  the variation 
of surface area of the 
droplet $\delS$ is chosen as the target energy, since $\gamma \delS$ indicates
the interfacial energy throughout the evolution: $\delS$ is zero only for a spherical droplet and is positive otherwise. The surface area is
 $S = 2\pi\int_{0}^{\pi}R^{2}\sin{\theta}\sqrt{1+\left[(1/R)(\partial R/\partial \theta)\right]^{2}}\d \theta$.
Assuming small deformation and thus $\frac{1}{R}\frac{\partial R}{\partial \theta} \ll 1$, a Taylor 
expansion yields
\begin{equation}\label{equ:taylor_area}
 S  =  2\pi\int_{0}^{\pi}R^{2}\sin{\theta}\lp 1+\frac{1}{2R^{2}}\lp\frac{\partial R}{\partial 
\theta}\rp^{2}\rp \d\theta.
\end{equation}
Plugging~\ref{equ:R_drop} into~\ref{equ:taylor_area}, the area variation $\Delta S = S-4 \pi$ is found to be
\begin{align}\label{equ:dels_linear}
 \delS/\lp 2\pi \delta^{2} \rp =\f^{\mathrm{T}} \Marea \f + o\lp \delta^{2} \rp, 
\end{align}
	where $\Marea$ is the so-called weight 
matrix~\citep{schmid2001tools} of size $m \times m$, with entries 
\begin{equation}
 \Marea\lp i,j \rp = 2\delta^{\mathrm{K}}_{ij} \lp 2i+1 \rp + \frac{1}{2}\lp 2i+1 \rp \lp 2j+1 \rp
\int_{0}^{\pi} P^{\prime}_{i}\lp \cos\t \rp P^{\prime}_{j} \lp \cos \t \rp \sin^{3} \t  \d\t.
\end{equation}
The optimal growth of $\delS$ can now be defined as
\begin{align}
\gareamax \lp T \rp = \max_{\f\lp 0 \rp} \left [G_{\Delta S} \lp T \rp 
=  \frac{\sqrt{\delS\lp T \rp} }{\sqrt{\delS \lp 0 \rp}} \right] = \max_{\f\lp 0 \rp} \left [G_{\Delta S} \lp T \rp 
=  \frac{\sqrt{\f^{\mathrm{T}} \Marea \f}} {\sqrt{{\f}^{\mathrm{T}}\lp 0 \rp \Marea \f\lp 0 \rp}} \right].
\end{align}
By Cholesky decomposition $\Marea = {\F}^{\mathrm{T}}\F$, the above equation is 
formulated as
\begin{align}
 \gareamax \lp T \rp = \max_{\f\lp 0 \rp} \left [G_{\Delta S} \lp T \rp 
 =  \frac{||\F \f\lp T \rp||_{2}} {||\F \f\lp 0 \rp||_{2}} \right].
\end{align}

In a similar way to how the asymptotic stability ($t \rightarrow \infty$) is determined by the eigenvalues of the evolution operator $\A$, the maximum instantaneous
growth rate of the perturbation energy  at $t = 0^{+}$ can be determined algebraically, expanding the matrix exponential $\exp(t\A) \approx I+t\A$ at $t=0^+$.
The growth rate of the excess area $\Delta S$ is then
\begin{equation}\label{eq:area_growth}
 \frac{1}{\Delta S}\frac{ d \Delta S}{d t} \bigg |_{t=0^+}  = \frac{\f^{\T}(0) \left[ \A^{\T} \F^{\T} \F +  \F^{\T} \F \A\right] \f(0)}{\f^{\T}(0) \F^{\T} \F \f(0)}.
\end{equation}
By introducing $\acca = \F \f(0)$, the maximum growth rate of $\Delta S$ is formulated as 
\begin{equation}\label{eq:area_growth2}
\max \frac{1}{\Delta S}\frac{ d \Delta S}{d t} \bigg |_{t=0^+}  = \max_{\acca} \frac{\acca^{\T} \left[ \F\A\F^{-1} + \lp \F\A\F^{-1} \rp^{\mathrm{T}} \right ] \acca}{\acca^{\T} \acca},
\end{equation}
which becomes the optimization of a Rayleigh quotient with respect to $\acca$. Because $\F\A\F^{-1} + \lp \F\A\F^{-1} \rp^{\mathrm{T}} $ is a symmetric operator,
the maximum is given by its largest eigenvalue,
\begin{eqnarray}\label{eq:weightnumrange}
\max\frac{1}{\sqrt{\delS} }\frac{d\sqrt{\delS}}{dt} \bigg|_{t=0^{+}}= s_{\mathrm{max}} \left[ 
\frac{1}{2}\lp \F\A\F^{-1} + \lp \F\A\F^{-1} \rp^{\mathrm{T}}
\rp \right],
\end{eqnarray}

where $s_{\mathrm{max}}\left[ \cdot \right]$ denotes the largest eigenvalue. 
This maximum instantaneous growth rate is
 commonly called the numerical abscissa
\citep{trefethen2005spectra}, which is closely linked to the numerical 
range $W_{\delS}\lp \A, \F \rp $  defined 
as the set of all Rayleigh  quotients,
\begin{equation}\label{eq:weightnumrange}
 W_{\delS}\lp \A, \F \rp  \equiv \{ z: z = \lp \F\A\F^{-1} \bp, \bp \rp / \lp \bp, \bp \rp \}.
\end{equation}
The numerical range is the convex hull of the spectrum for a normal operator 
(and is therefore always in the stable half plane 
$z_{r}<0$ for a stable operator) , but can extend significantly  to even 
protrude into the unstable half-plane $z_{r}>0$ for stable non-normal 
operators. Its 
maximum protrusion is equal to the numerical abscissa and thus determines the 
maximum energy growth rate at $t = 0^{+}$.

\section{Non-modal analysis: results}\label{sec:linear_nonmodal}
\subsection{Transient growth and numerical range}
\begin{figure}
    \hspace{-0.5em}\includegraphics[scale = 0.44] {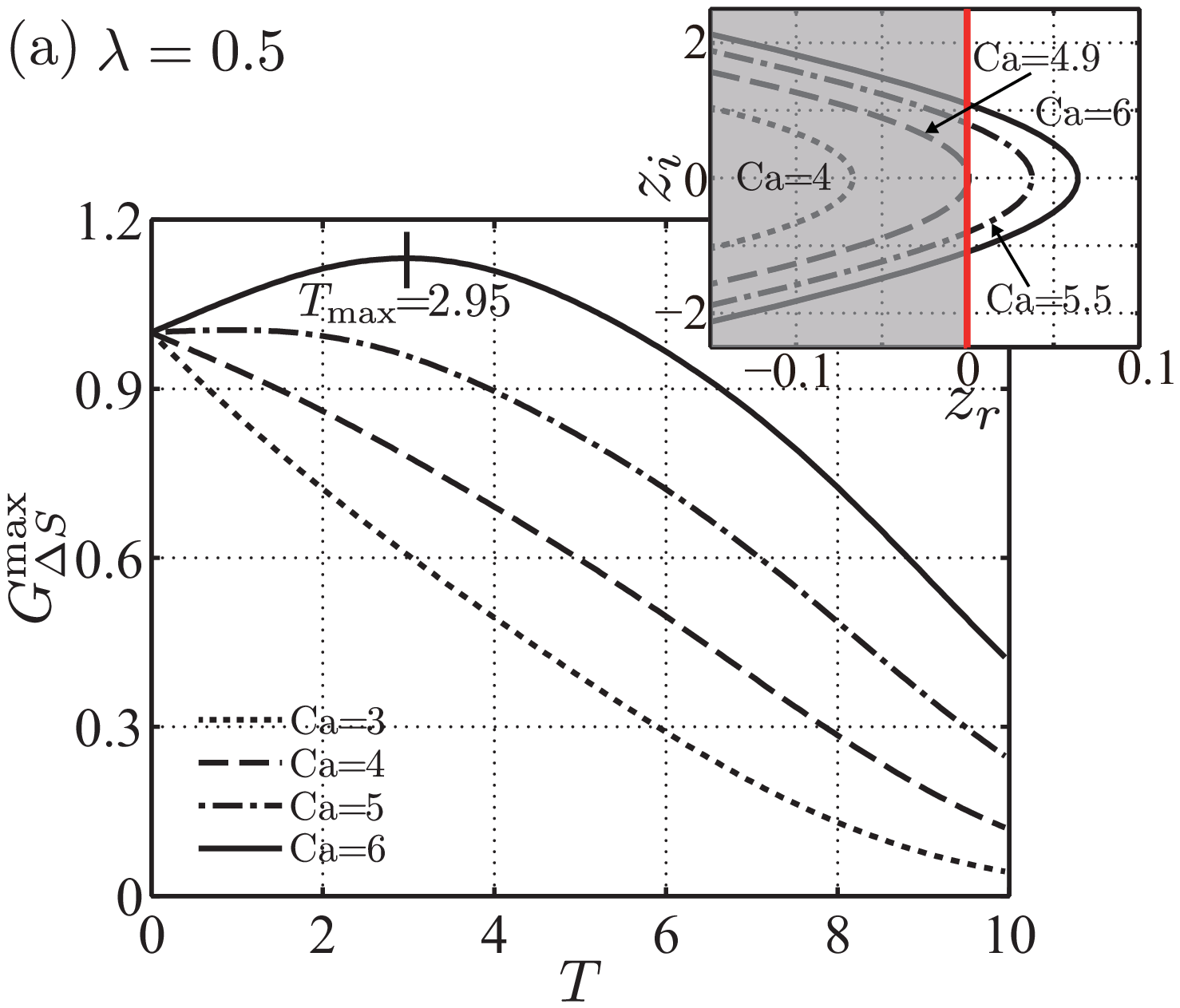}
    \hspace{1em}\includegraphics[scale = 0.44] {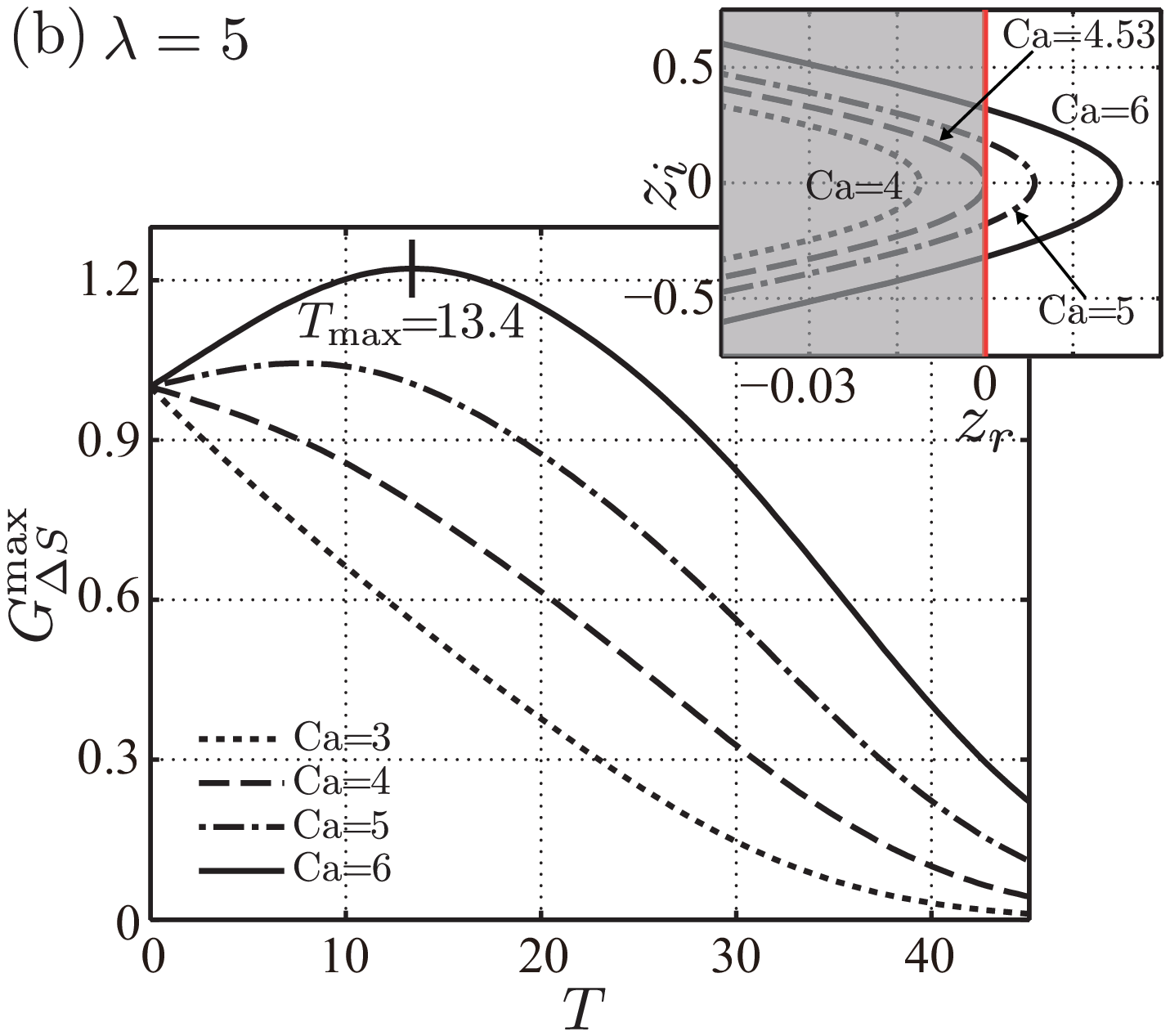}
       \caption{
       The optimal growth of the interfacial energy $\gtareamax$ versus the 
nondimensional time $T$, for viscosity ratio $\lambda=0.5$ (a) and 
$\lambda=5$ (b); for each case, four capillary number $\Ca$s are shown and
for the highest Ca, the time $\tmax$ corresponding to the peak energy growth is marked. The inset
shows the boundary of the numerical range $\lp z_{r}, z_{i}\rp$.
}
\label{fig:gtarea}
\end{figure}

In figure ~\ref{fig:gtarea}, we show the optimal growth of the interfacial energy $\gtareamax$ for viscosity 
of ratios $\lambda=0.5$ and $5$, varying the capillary number $\Ca$.
 The threshold value of $\Ca$ to yield transient growth is between $4$ and $5$, in accordance with
the rightmost boundary of the numerical range (see inset) depicted in the complex plane $\lp z_{r}, z_{i} \rp$. The boundary is
almost tangent to $z_{r}=0$ at $\Ca \approx 4.9$ for $\lambda=0.5$ and $\Ca 
\approx 4.53$  for $\lambda=5$ 
representing the  critical capillary number $\Cacri$ above which the maximum 
energy growth rate at $t=0$, $\max_{\f\lp 
0 
\rp}\frac{1}{\sqrt{\delS} }\frac{d\sqrt{\delS}}{dt}|_{t=0^{+}}$, is positive, guaranteeing transient growth.

\subsection{Linear growth and shape evolutions}
\begin{figure}
\centering
	\includegraphics[scale=0.33]{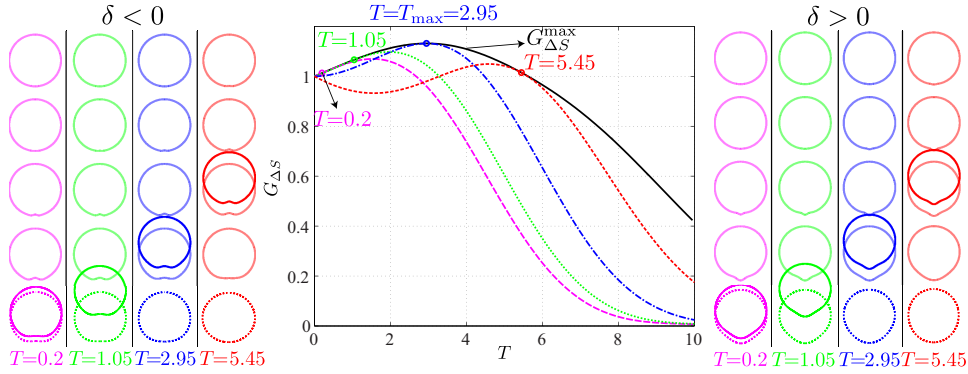}
		\caption{Linear growth $\garea$ of the interfacial energy of the droplets with an optimal initial 
perturbation $\fo \lp 0 \rp$ for the target times $T=0.2$, $1.05$, $2.95$ and $5.45$; the solid curve indicates the 
optimal growth 
$\garea^{\mathrm{max}}\lp T \rp$ and it reaches its peak at $T=T_{\mathrm{max}}=2.95$. The linear shape evolution of 
the 
perturbations are shown for negative and positive $\delta$, on the left and right panel respectively.
}  
\label{fig:linear_evol_four_opt}
\end{figure}

Non-modal analysis not only predicts the maximum energy growth over a particular time interval,
but also provides the optimal perturbation, i.e. the initial shape coefficients $\foi$ 
that ensure the optimal gain at horizon $T$.
Figure~\ref{fig:linear_evol_four_opt} depicts the individual energy 
gains $\garea$ 
for four optimal initial conditions $\foi$ 
corresponding to $T=0.2$, $1.05$, $3.95$ 
and $5.45$, with $\lambda=0.5$ and $\Ca = 6 $. 
Their gain profiles are tangent to 
$\gtareamax$ at $t=T$. The optimal perturbation targeting $T=\tmax=2.95$ coincides with
the optimal growth $\gdelsmax$ at its peak.

Assuming small deformation amplitude and integrating equ.~\ref{equ:linear_evol} 
in time, the linear shape evolution is 
readily reconstructed  for the droplets with the four optimal initial 
conditions,  depicted in 
figure~\ref{fig:linear_evol_four_opt}, at time $t=0$ (dashed), 
 $2.5, 5, 7.5, 10$ (light solid) and the target time $t=T$ (solid); the evolution is shown for 
negative/positive $\delta$
in (a)/(c). For both signs, the initial perturbation is mainly introduced near the tail ($\theta=0$) of 
the droplet where the interface is respectively flattened for $\delta<0$ and stretched for $\delta>0$
while the front part of 
the droplet remains spherical. In accordance with the modal analysis implying a linearly stable evolution, the
perturbations eventually decay and the droplets finally recover a 
spherical shape.

\section{Nonlinear analysis}~\label{sec:lin_non_evol}
\subsection{Nonlinear energy growth and shape evolution using DNS}
As the droplets deform more and more on increasing the initial perturbation amplitude,
nonlinearities become significant and the droplet evolution cannot be adequately described by 
the linearised equations. We resort to DNS to address the non-linear dynamics using a 
three-dimensional axisymmetric boundary integral implementation, following the standard approach of ~\citet{koh1989stability}.

\begin{figure}
        \centering
    \subfigure[Energy growth $\gtarea$]{\label{fig:nonlinear_gt}
    \hspace{-3em}\includegraphics[scale = 0.38] {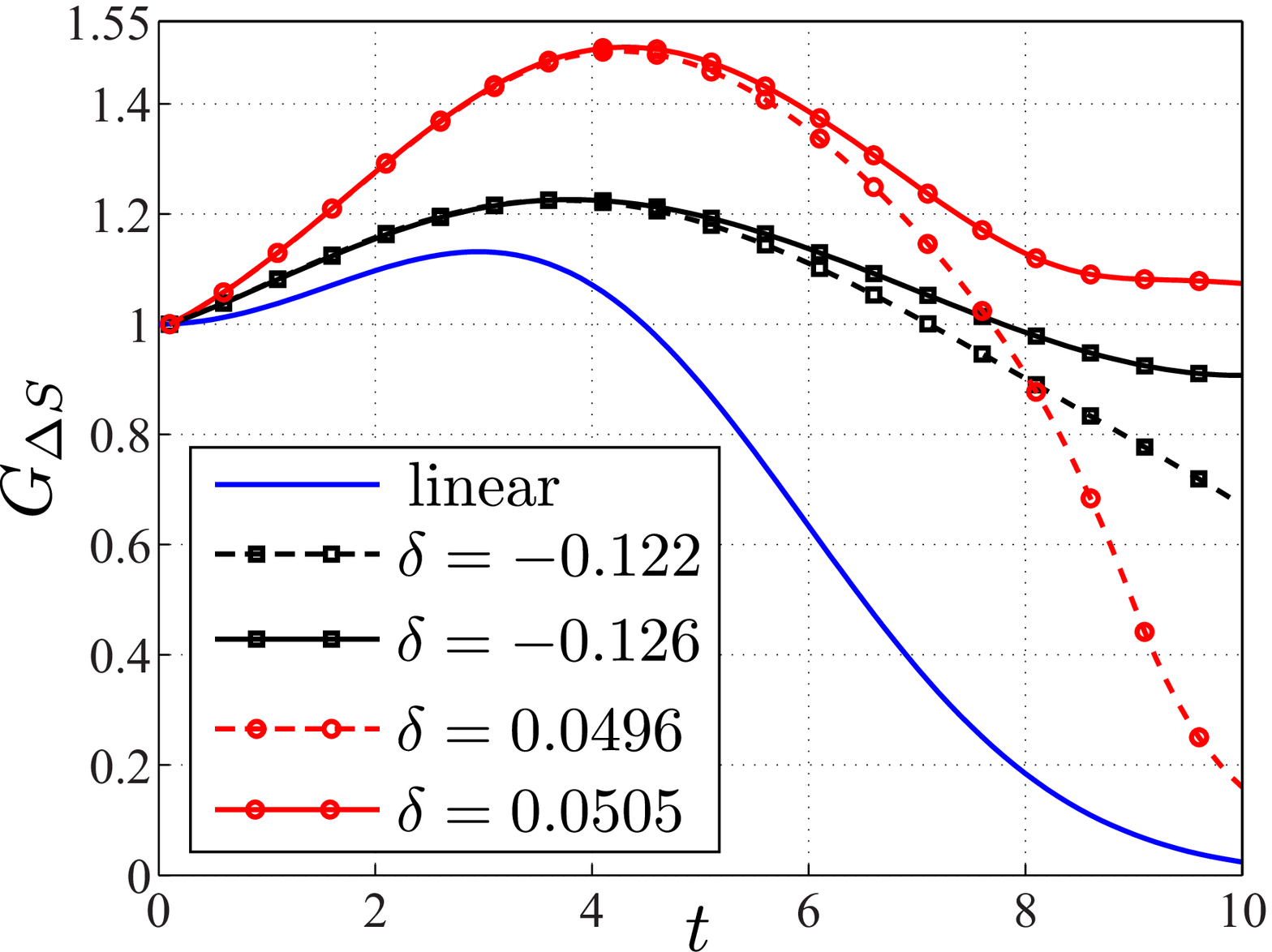}
 }
    \subfigure[Shape evolution]{\label{fig:nonlinear_shape}
    \hspace{0em}\includegraphics[scale = 0.45] {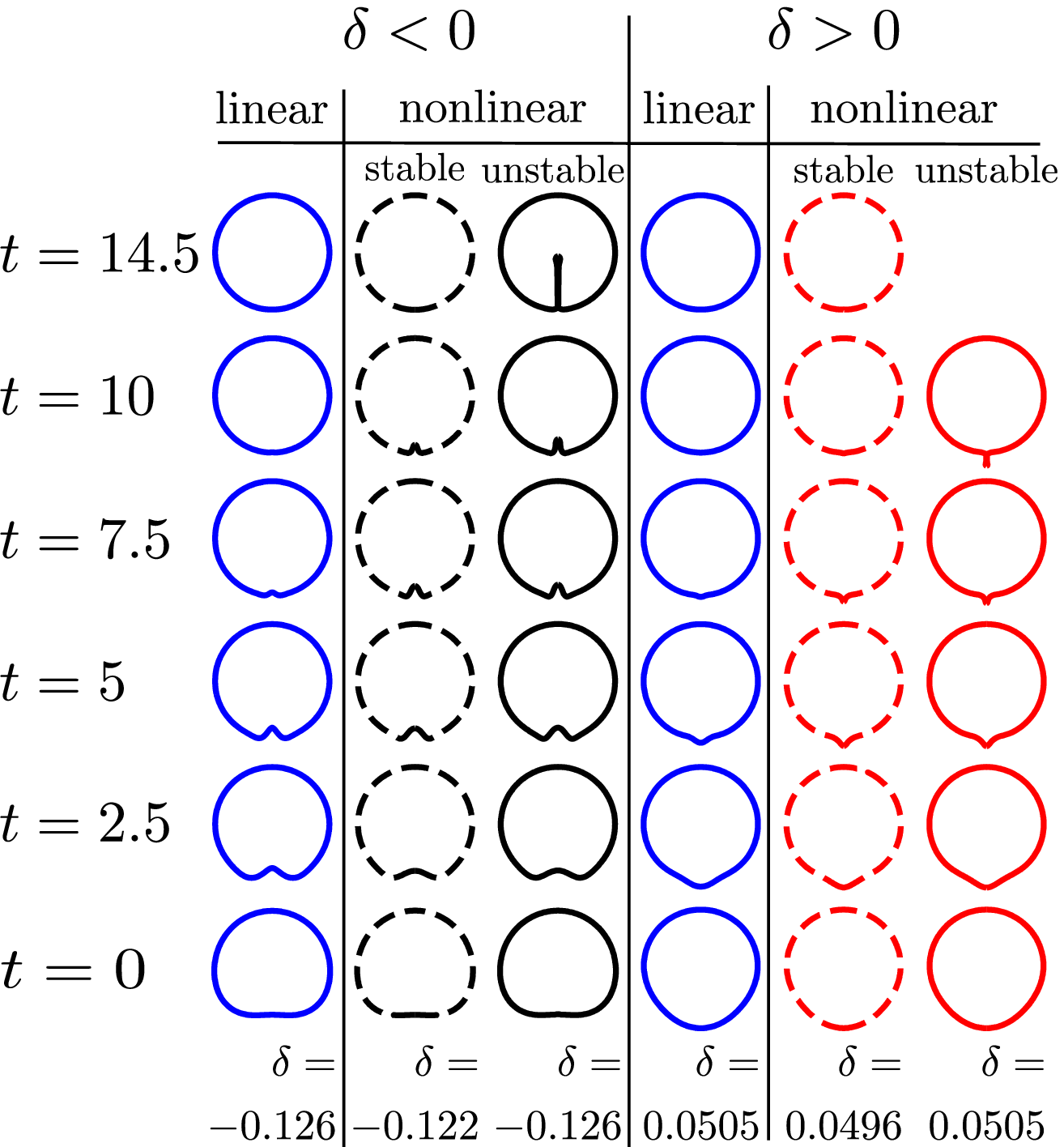}
 }
       \caption{~\subref{fig:nonlinear_gt}: Nonlinear energy growth $\garea$ of the droplets with the 
optimal perturbation $\foi$  for the target time $T=\tmax=2.95$; the solid curve indicates the linear  energy 
growth.
For positive $\delta$, $\garea$ of droplets with $\delta=0.0496$ and $0.0505$ are shown, the 
former/latter being stable/unstable; for negative $\delta$, the chosen value leading to stable and unstable evolution 
is $\delta=-0.122$ and $\delta=-0.126$ respectively. 
~\subref{fig:nonlinear_shape}: The shape evolutions of the 
corresponding droplets.}
\label{fig:nonlinear_lam05_ca6}
\end{figure}

We focus on the droplets of $\lambda=0.5$ and $\Ca  = 6  $ with the optimal perturbation $\fomax \lp 0 \rp$ 
 achieving the peak of the optimal energy growth $\gdelsmax$  at $\tmax$. Two slightly different magnitudes of perturbation
$\delta=0.0496,0.0505$ are chosen for the positive $\delta$ and similarly $\delta=-0.122, 
-0.126$ for the negative case. Their energy growth $\gtarea = \sqrt{\frac{\delS \lp 
t \rp}{\delS\lp 0 \rp}} $ is plotted in figure~\ref{fig:nonlinear_gt}, 
together with the linear counterpart $\gtarea$ using 
equ.~\ref{equ:dels_linear}. The linear and non-linear energy growth share the 
same trend in the initial growing phase $t<3$, but differ as the former is 
approximated by a truncated Taylor expansion.
For the two values of $\delta$ with the same sign but slightly different 
magnitude, the energy growth curves almost 
collapse before reaching their peaks at $t \approx 4$, but diverge afterwards; 
$\garea$ decays for the smaller magnitudes $\delta = 
-0.122$ and $\delta = 0.0496$ indicating stable evolutions but maintains a 
sustained value around $1$ for larger 
initial amplitudes $\delta = -0.126$ and $\delta =  0.0505$, implying the onset 
of instability.

The shape evolutions of  droplets are shown in 
figure~\ref{fig:nonlinear_shape}. For $\delta=-0.122$ and 
$-0.126$, no significant difference  is observed for $0<t<7.5$: an inward 
cavity develops at the rear and sharpens; it is subsequently smoothed out and 
disappears for $\delta=-0.122$ while it
keeps growing to form a long indentation for $\delta=-0.126$. These two values 
of $\delta$ bound  a threshold initial amplitude required to excite  nonlinear 
instabilities.  A similar trend is  found for positive values of $\delta$, while the 
instability arises through the formation of a dripping tail. 

It becomes natural to introduce $\delcri$, the critical magnitude of 
the perturbation above/below which
the evolution of the drop is unstable/stable. Parametric computations are conducted to identify
$\delcri^{\pm}$ within a confidence interval (for instance $ \delcri^+ \in \left[ 0.0496,  0.0505 
\right]$ and  $ \delcri^- \in \left[ -0.122, -0.126  \right]$ as in figure~\ref{fig:nonlinear_gt}). Searching in 
both directions,
the critical amplitude is then defined as $\delcri=\mathrm{min}(|\delcri^{+}|,|\delcri^{-}|)$.
When $\lambda=0.5$ and $\Ca=6$, $|\delcri^{-}|>|\delcri^{+}|$, implying that 
the instability tends to favour an initially stretched tail with respect to a 
flattened bottom; otherwise when $\lambda=5$ the situation reverses  $\left( |\delcri^{-}|<|\delcri^{+}| \right)$, as discussed in next section. 

\subsection{Critical amplitude of the perturbation $\delcri$}
\begin{figure}
        \centering
    \subfigure[$\lp\lambda,\Ca\rp=\lp0.5,6\rp$]{\label{fig:critdelta_lam05}
    \hspace{0em}\includegraphics[scale = 0.27]  {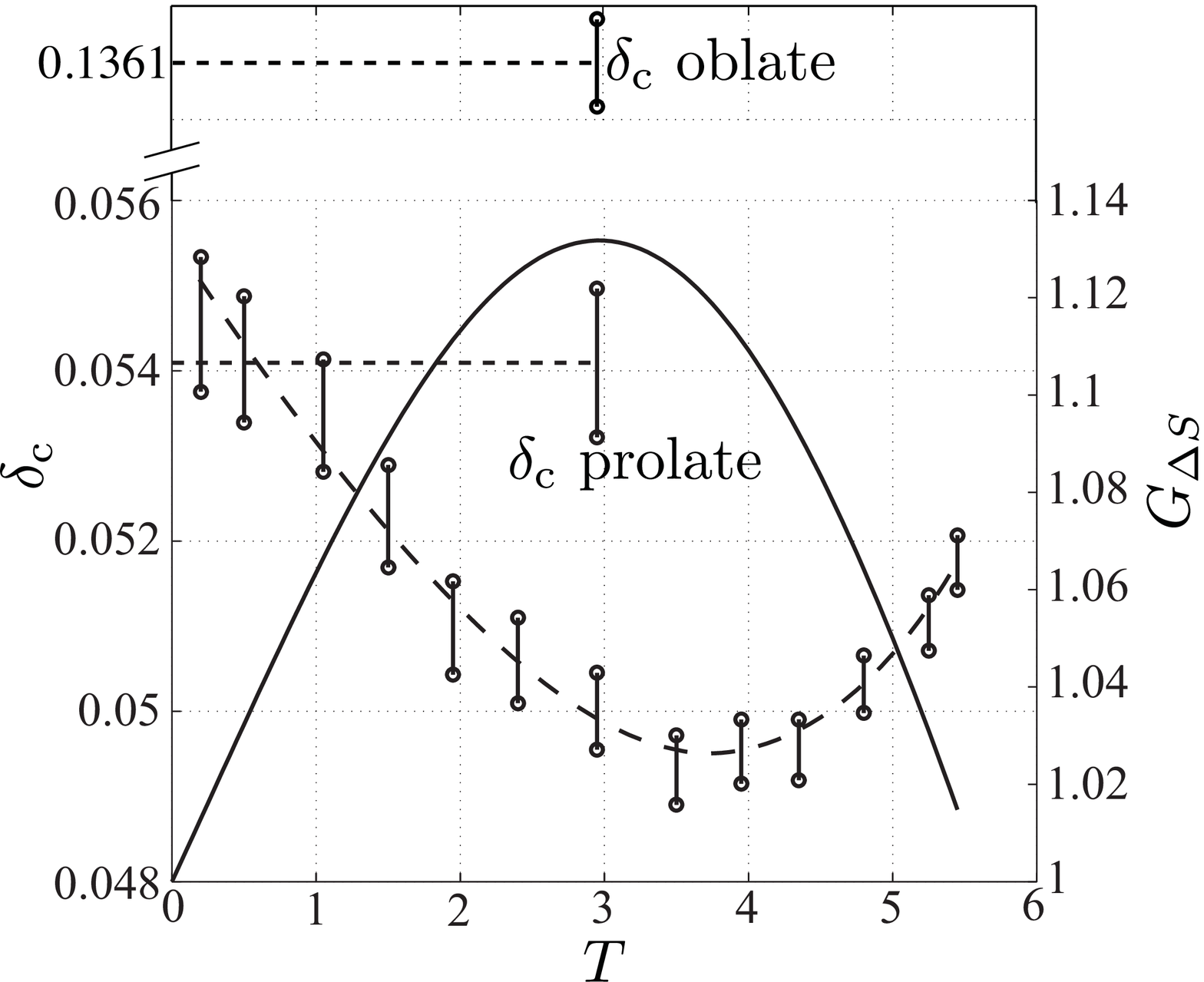}
 }
    \subfigure[$\lp\lambda,\Ca\rp=\lp5,6\rp$]{\label{fig:critdelta_lam5}
    \hspace{0em}\includegraphics[scale = 0.27]  {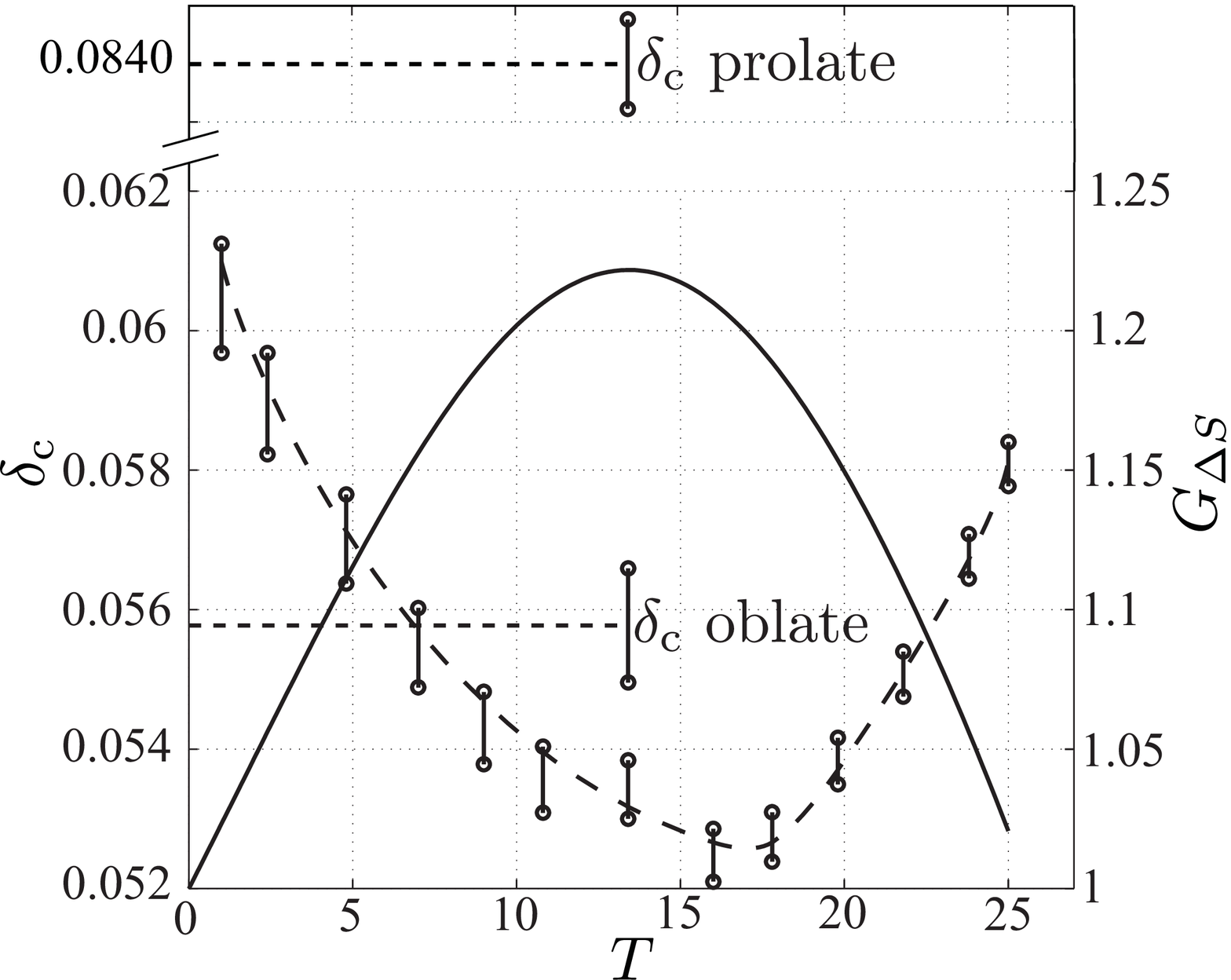}
 }
       \caption{The critical perturbation magnitude $\delcri$ for:  
~\subref{fig:critdelta_lam05}: $\lp \lambda,\Ca \rp = \lp0.5, 6\rp$ and
~\subref{fig:critdelta_lam5}: $\lp \lambda, \Ca \rp = \lp5, 6\rp$. The upper and lower limits of $\delcri$ (measured by 
the left scale) are plotted versus the target time $T$, with a curve fitted to show the trend.
Accordingly, the linear 
energy growth $\garea$ (measured by the right scale) is provided. ${\delta}^{\mathrm{P}}_{\mathrm{c}}$ and  
${\delta}^{\mathrm{O}}_{\mathrm{c}}$ is the 
critical magnitude for an initially prolate and oblate respectively.
}
\label{fig:critdelta}
\end{figure}

Following the description of the previous paragraph, 
the critical deformation amplitude $\delcri \lp T \rp$  
can be determined for any targeting time $T$ and associated optimal initial 
perturbation $\fo \lp 0 \rp$. The critical deformation amplitude $\delcri \lp T \rp$ is plotted in 
figure~\ref{fig:critdelta} for  $\Ca =  6$ and  both $\lambda=0.5$ and 
$\lambda=5$,
together with the optimal growth $\gareamax$. 
The critical deformation amplitude $\delcri $ is negatively correlated with 
$\gareamax$ and 
corresponds to a target time $T$ slightly larger than 
$\tmax$ where the peak transient growth is reached. This shows that the 
transient growth 
reduces the threshold non-linearity needed to 
trigger instabilities and consequently the critical magnitude of the initial 
perturbation.

We also determined the critical amplitude $\delcri^{\mathrm{P}/\mathrm{O}}$ for 
an initially prolate ($\mathrm{P}$) / oblate ($\mathrm{O}$) ellipsoidal droplet 
to be unstable, as reported 
in figure~\ref{fig:critdelta}.  
When the fluid inside the droplet is less viscous than the one 
outside, i.e. $\lambda<1$, an initially prolate droplet is more unstable,  $\delcri^{\mathrm{P}}$ is 
less than half that of an oblate droplet $\delcri^{\mathrm{O}}$; the trend reverses as $\lambda>1$. 
Such an observation is in 
agreement with the results of \citet{koh1989stability} using DNS (see fig. 11 
of their paper). 
As expected, the minimum $\delcri$ using the optimal perturbations is smaller than 
$\mathrm{min}(\delcri^{\mathrm{P}},\delcri^{\mathrm{O}})$  based on the limited
family of ellipsoidal shapes.

\begin{figure}
    \hspace{-0.5em}\includegraphics[scale = 0.35] {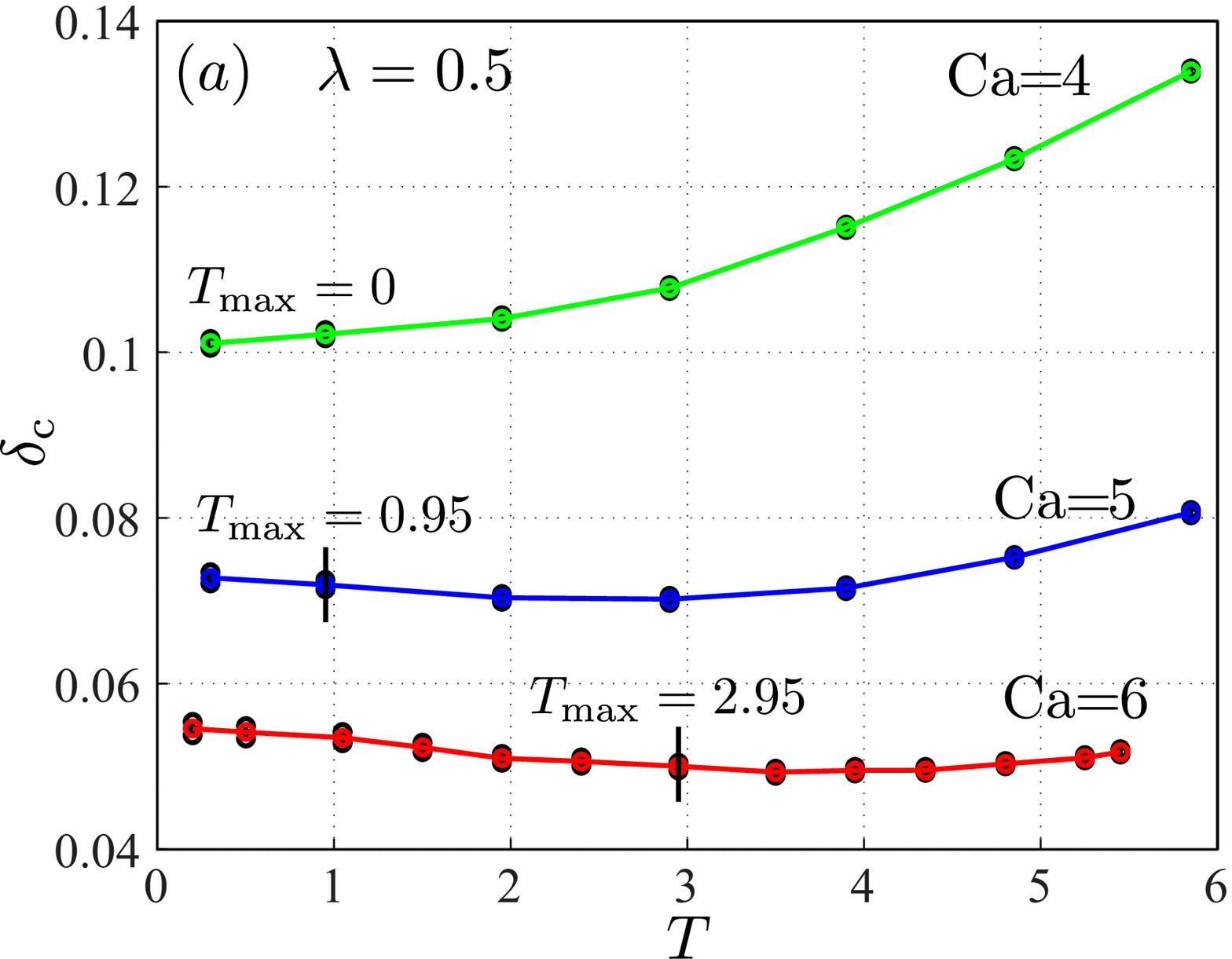}
    \hspace{2em}\includegraphics[scale = 0.35] {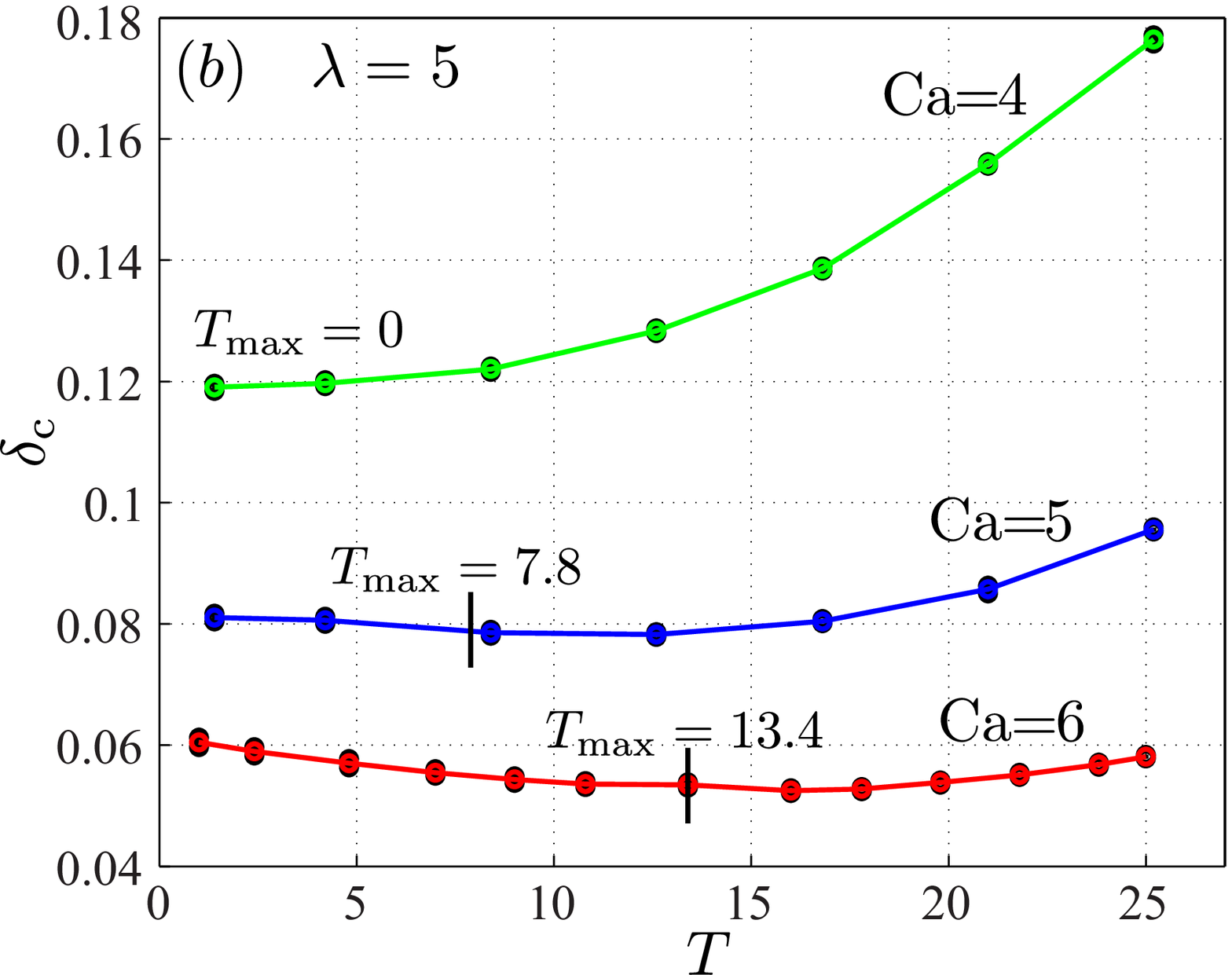}
       \caption{Akin to figure~\ref{fig:critdelta},  adding 
$\delcri$ of two smaller $\Ca$s for $(a)$: $\lambda=0.5$ and $(b)$: $\lambda=5$.}
\label{fig:critdelta_ca}
\end{figure}

So far, we have analysed the critical amplitude $\delcri$ of perturbations 
exhibiting transient energy growth. 
We would 
like to know how it varies as the transient growth decreases
and even disappears as it is
suppressed by high surface tension. In addition to $\Ca=6$, the 
time-dependence of $\delcri$  is shown in figure~\ref{fig:critdelta_ca} for 
$\Ca=4, 5$. As expected, $\delcri$ 
increases
with decreasing $\Ca$, by a factor of approximately $3$, varying from the highest to the 
lowest $\Ca$. 
With respect to $T$, $\delcri$ varies non-monotonically for $\Ca=5, 6$ showing 
transient growth. In the absence of transient growth, like for $\Ca=4$, 
$\delcri$ increases with $T$ monotonically. Indeed, without transient 
growth, the energy decays monotonically and $\tmax=0$, hence the minimum 
$\delcri$ appears at $T \approx 0$.

\section{Conclusion and discussions}\label{sec:conclusions}
\begin{figure}
        \centering
    \hspace{0em}\includegraphics[scale = 0.37]  {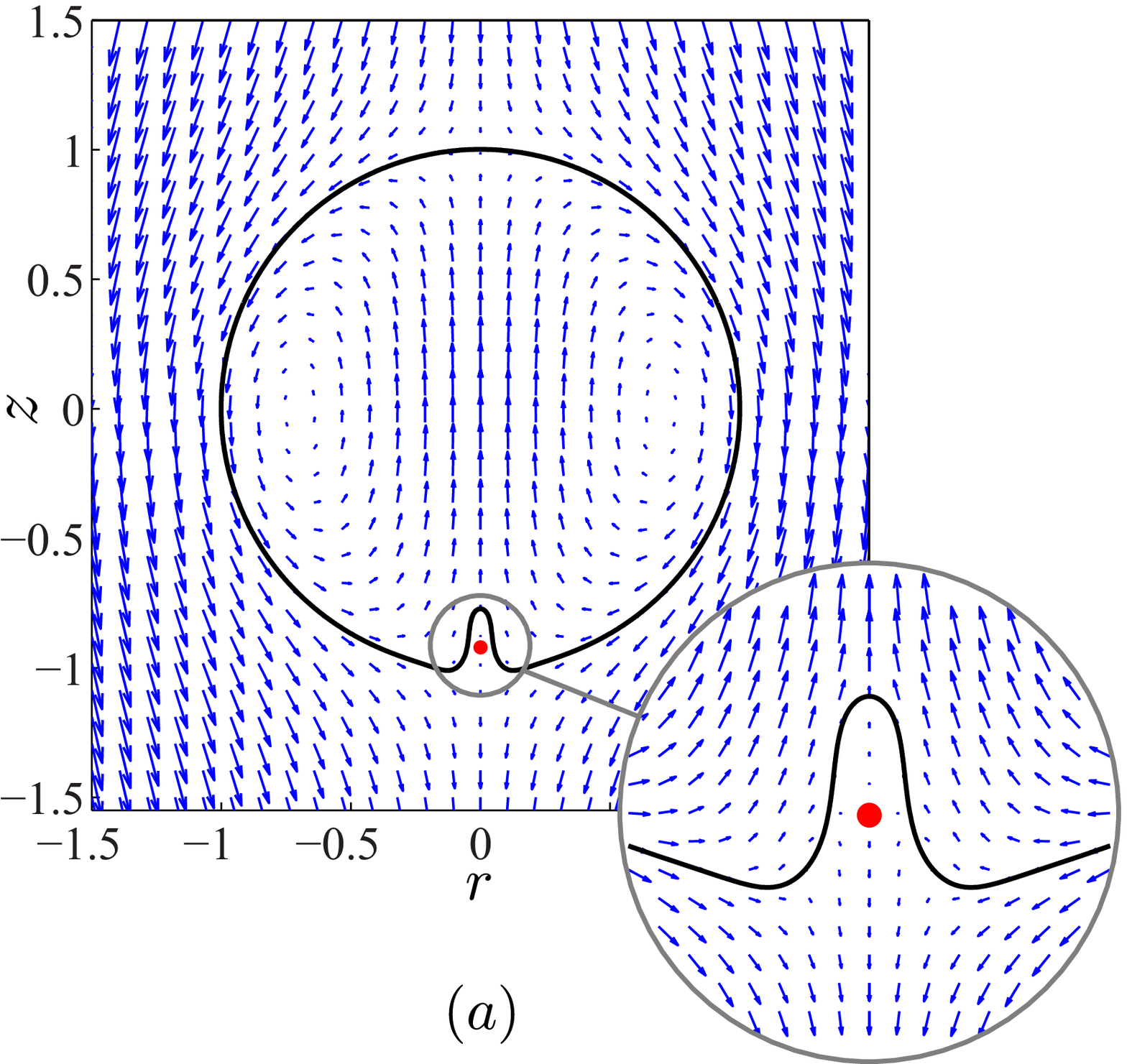}
    \hspace{0em}\includegraphics[scale = 0.37]  {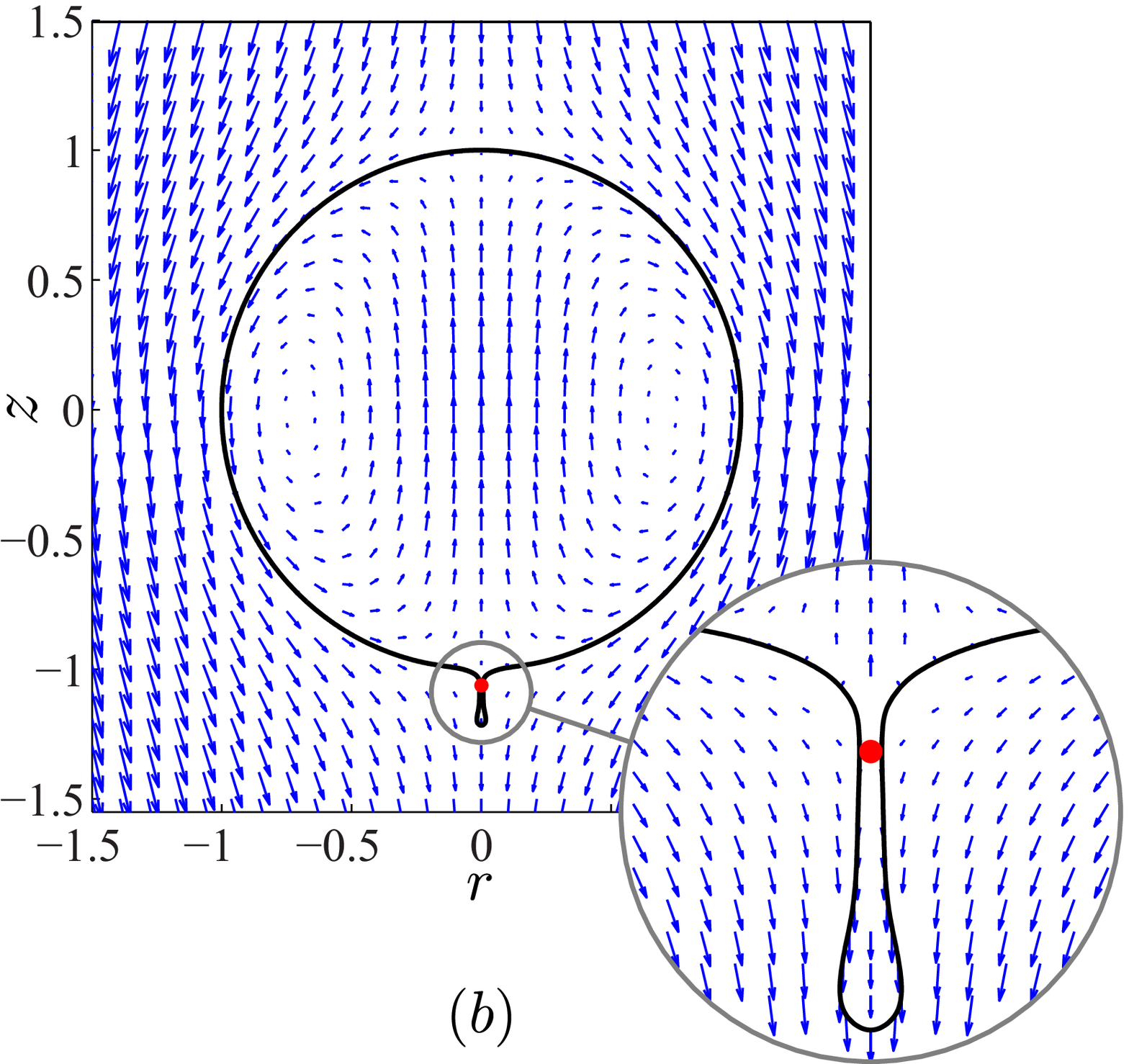}
       \caption{The flow field co-moving with the droplet,
       using the optimal initial coefficient $\fomax \lp 0 \rp$ ,
       when $\lp\lambda,\Ca\rp=\lp0.5,6\rp$, $(a)$: $\delta=-0.126$ and 
$(b)$: $\delta=0.0505$. 
The red dot indicates the stagnation point of the flow.}
\label{fig:flowfield}
\end{figure}
In this paper, we have performed non-modal analysis and DNS to investigate the 
shape instabilities of an intertialess rising droplet which 
tends to recover the spherical shape, the attractor solution, due to surface 
tension. 
For sufficiently low surface tension, transient 
growth of the interfacial energy arises and leads to a bypass transition. This reduces the initial 
disturbance amplitude required to trigger instability, 
hence significantly decreasing the 
threshold magnitude of perturbation for the droplet to escape the basin of 
attraction. This magnitude is negatively correlated to the optimal 
growth of the interfacial energy.

We now compare our results with the work of~\citet{koh1989stability} who employed DNS 
to identify 
the critical capillary number $\Cacri$ leading to shape instabilities of an 
initially prolate or oblate ellipsoidal sedimenting droplet;
the magnitude of  perturbation is $\Delta=\frac{L-B}{L+B}$ (see 
figure~\ref{fig:sketch}). For their lowest 
magnitude $|\Delta|=1/21$ considered, $\Cacri \in (4, 5)$ for $\lambda=0.1, 
0.5$ and $5$, 
indeed close to our prediction: $\Cacri \approx 5.42$, $4.9$ and $4.53$ 
respectively for 
the same $\lambda$. 
Additionally, ~\citet{koh1989stability} observed that for a viscosity ratio $\lambda<1$/$\lambda>1$, the first unstable 
pattern appears as a protrusion/indentation developing near the tail of the droplet
that is initially a prolate/oblate. The trend holds in our case even though we search over all possibilities for the 
most 'dangerous' initial perturbation instead of using an initially 
ellipsoidal shape. This is also reflected from the initial shapes: as $\lambda<1$/$\lambda>1$, the optimal shape shares 
a common feature with an oblate/prolate ellipsoid, namely its rear interface is compressed/stretched. 

To explain the dependence of the instability patterns on the viscosity ratio 
$\lambda$, let us
focus on the velocity field near the tail of the droplet (see 
figure~\ref{fig:flowfield}), where the flow resembles a uniaxial 
extensional flow, drawing the tip into the drop on the top side and pulling it 
outwards on the other side. We suggest that this 
imbalance induces the onset of the shape instability. The internal (respectively 
external) viscous force on the 
tip is $\mu_{1}  \partial u^{\mathrm{tip}}_{z}/\partial z $ (respectively $\mu_{2} 
\partial u^{\mathrm{tip}}_{z} / \partial z$). When
$\mu_{1}<\mu_{2}$, i.e. $\lambda<1$, the external viscous effect overcomes the internal one, hence the perturbation 
tends to be stretched outward to form a protrusion; otherwise, when $\lambda>1$, 
it is prone to be sucked inwards to form an indentation.

Developed originally for hydrodynamic stability analysis, non-modal tools 
have here demonstrated the predictive 
capacity for the inertialess shape instabilities of capillary interfaces. 
This work
might stimulate the application of non-modal analysis for 
complex  multiphase flow instabilities even at low Reynolds number.

\section*{Acknowledgements}
L.Z. thanks Francesco Viola for the helpful discussions. We thank the anonymous referee for pointing 
out an incorrect coefficient in our previous derivation. Computer time from 
SCITAS at EPFL is acknowledged, and the European Research Council is acknowledged 
for funding the work through a starting grant (ERC SimCoMiCs 280117).

 \bibliographystyle{jfm}

\end{document}